\pdfoutput=1

\documentclass{easychair}

\usepackage{doc}
\usepackage{makeidx}

\begin{document}

\title{IPv6 Prefix Alteration: An Opportunity to Improve Online Privacy\footnote{This paper was presented at the 1st Workshop on Privacy and Data Protection Technology (PDPT 2012), co-located with the Amsterdam Privacy Conference (APC 2012), October 9, 2012.}}

\titlerunning{IPv6 Prefix Alteration}

\author{
    Dominik Herrmann
\and
    Christine Arndt
\and
   Hannes Federrath
}

\institute{
  Department of Informatics, University of Hamburg, Germany\\
  \email{\{herrmann,9arndt,federrath\}@informatik.uni-hamburg.de}
 }

\authorrunning{Herrmann, Arndt, Federrath}

\clearpage

\maketitle

\begin{abstract}
This paper is focused on privacy issues related to the prefix part of IPv6 addresses. Long-lived prefixes may introduce additional tracking opportunities for communication partners and third parties. We outline a number of prefix alteration schemes that may be deployed to maintain the unlinkability of users' activities. While none of the schemes will solve all privacy problems on the Internet on their own, we argue that the development of practical prefix alteration techniques constitutes a worthwile avenue to pursue: They would allow Internet Service Providers to increase the attainable privacy level well above the status quo in today's IPv4 networks.
\end{abstract}

\pagestyle{empty}

\section{Introduction}

Many Internet service providers (ISPs) throughout the world are now in the process of integrating IPv6 into their Internet access products for retail customers and corporate clients alike.
One of the most important features of the IPv6 protocol is its huge address space. This will enable users to assign a life-long stable, unique and globally routable IP address to each personal device. In principle, all users will be able to set up public services at home and be able to communicate from and to their devices from any point in the network. The resulting \emph{end-to-end reachability} is considered a major improvement over the current situation \cite{RFC5902}. While the initial IPv6 specification from 1995 does already contain a variety of security considerations (namely, message integrity protection as well as encryption based on IPsec), privacy aspects have not played a major role in the beginning \cite{RFC1883}.

\begin{figure}[htbp]
        \centering
                \includegraphics[scale=0.65]{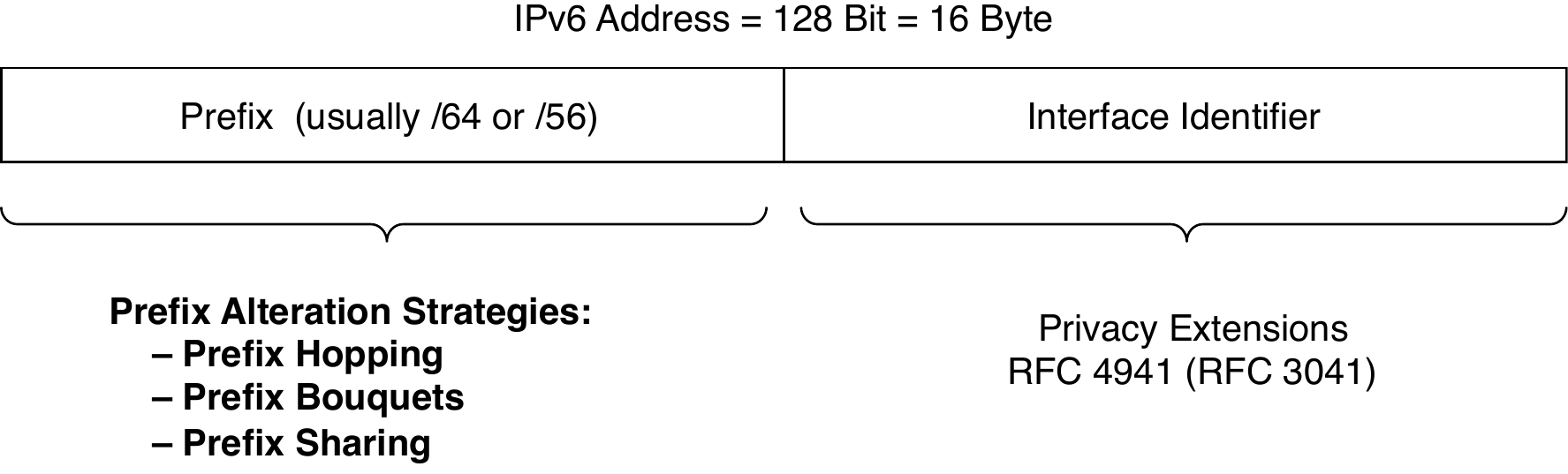}
        \caption{IPv6 Address Prefix and Interface Identifier}
        \label{fig:fig_ipv6_adr}
\end{figure}

The overall design of IPv6 and especially the property of end-to-end reachability are in conflict with the desirable privacy property of \emph{unlinkability}, i.\,e., an attacker should not be able to relate multiple activities of a user based on his or her (possibly long-lived) IP address (cf. \cite{PH09} for a more formal description of unlinkability). According to the initial specification of IPv6 \cite{RFC1883} network nodes derive the interface identifier part of an IPv6 address (one of the two parts of an IPv6 address, see Fig.~\ref{fig:fig_ipv6_adr}) from the Medium Access Control (MAC) address of the respective network adapter, which is determined at the time of manufacturing and does not change over time. As a result, every IPv6 node can be uniquely identified via its static interface identifier. This design has received much attention in the privacy community and was resolved with the specification of the so-called IPv6 Privacy Extensions \cite{RFC4941}. Once Privacy Extensions are enabled, a node will choose a random interface identifier, which is changed on a regular basis (e.\,g. once per day).

In this paper we want to draw the attention to another privacy issue, which has been often overlooked so far: Tracking the activities of Internet users may not only be possible because of long-lived interface identifiers, but also if the first part of their IPv6 addresses, the \textbf{prefix} that is assigned by the ISP to the customer, remains constant over time. In fact, according to initial press statements released so far (\cite{heise2011}, for instance), some ISPs, including German market leader Deutsche Telekom, appear to be planning to assign long-lived prefixes to the customers of residential broadband products by default. This move would effectively decrease the privacy of many users, who are connected to the Internet with dynamic IPv4 addresses that change on a daily basis today.

Long-lived IPv6 prefixes would introduce a tracking mechanism, which could be used in addition to HTTP cookies. With tracking cookies ad networks such as Doubleclick or Google Analytics are already collecting sizeable amounts of user data for the purpose of profiling today. Long-lived IPv6 prefixes would allow them to maintain linkability even if users deleted all cookies in their browser. This is different from the current situation because tracking solely based on IPv4 addresses is usually infeasible: First of all, a large number of users is connected to the Internet using dynamic IPv4 addresses that change frequently, and secondly, multiple users may share a single IP address at a given point in time using Network Address Translation (NAT) or proxies.

With the growing awareness and driven by the recent ``Do-not-Track'' (cf.~\cite{dnt2011}) discussions we expect that an increasing number of users will eventually understand how tracking works and how it can be prevented by Privacy Enhancing Technologies. Instead of introducing new tracking possibilities with IPv6, ISPs could make use of the huge address space of IPv6 in order to offer improved privacy to their customers. This chance was already mentioned by the German IPv6 Council \cite{leitlinien-IPv6} and should now be discussed further.

In this paper we propose three design options relating to the assignment of IPv6 address prefixes to customers, which could be implemented by ISPs to protect the unlinkability of their users' actions against third parties. Our techniques consist of frequently changing address prefixes, distributing concurrent traffic across multiple prefixes and sharing prefixes among multiple users.

\section{Related Work}
\label{sec:related}

Sakurai et al.~\cite{Sakurai:07} propose a mechanism that allows two nodes to communicate privately by switching through a list of previously negotiated IPv6 interface identifiers. The objective of their technique is to provide relationship anonymity against eavesdroppers. To this end the nodes establish a secret key (possibly out-of-band), which is used to independently derive a shared list of randomised IP addresses in a deterministic way. During communication both nodes iterate over their address list in a synchronised manner and assign an ephemeral IP address to their respective network interfaces. This approach resembles \emph{frequency hopping} that is used in many wireless networks today.

Dunlop et al.~\cite{Dunlop:12} present the ``Moving Target IPv6 Defense'', another technique that dynamically obscures sender and receiver IPv6 addresses. The solution is quite similar to the one by Sakurai et al. as it replaces the interface identifier of sender and receiver addresses in a synchronised yet random manner using a pre-shared secret key. Additionaly, all packets are encapsulated in an encrypted tunnel to prevent payload analysis. The authors also include an empirical evaluation of their technique that demonstrates its practical feasibility.

The aforementioned approaches by Sakurai et al. and Dunlop et al. show that IPv6 offers a working foundation that can be used to implement privacy-enhancing techniques. While their proposals are of independent interest, they do not solve the privacy issues caused by long-lived prefixes we are interested in: First of all, they operate on the interface identifier only and completely neglect tracking via constant prefixes. Secondly, they require senders and receivers to negotiate a symmetric secret key before any communication can take place, which may scale poorly for the whole Internet. And finally, they protect only against tracking efforts by outsiders, while the two communication partners themselves are assumed to be trusting each other. Such a trust relationship does not exist in our scenario, in which users want to protect themselves against tracking by web sites or ad networks.

Lindqvist and Tapio \cite{Lindqvist:2008:PPP:1456403.1456416} propose to implement a \emph{protocol stack virtualisation} technique in the operating system of networked clients. Instead of only changing IP addresses, their solution takes into account that identifiers on other layers of the network stack can also be used to track users  (e.\,g., MAC addresses and port numbers). The design requires a locally installed translation daemon that replaces all identifiers on all layers at once for every new outbound flow of the client, which ensures that consecutive as well as concurrent activities remain unlinkable. A prototypical implementation of the system is shown to be compatible with several well-known Linux applications. As it solely operates on the client side, protocol stack virtualisation cannot influence the provider-controlled address prefixes, though.

The most relevant related work for our scenario has been published by Raghavan et al. \cite{Enlisting-ISPs}. The authors observe that all currently active customers of an ISP form an anonymity group, in which individual users can hide if addresses assigned to the customers are changed frequently. ISPs are supposed to assign two addresses to each customer: a \emph{hidden address} that is replaced with a randomly assigned address for every outbound flow from the customer's network as well as a \emph{sticky address} that remains constant over time to provide for reachability. The ISP runs a network address translation gateway that rewrites source and destination addresses on the fly. Although the technique is only described in the context of existing IPv4 networks it could be adapted to IPv6 prefixes. While Raghavan et al. present a single alternative, which resembles our Scheme 2, we focus on surveying and discussing various schemes that are conceivable in IPv6 networks in this paper.

\section{Unlinkability via Prefix Alteration}
\label{sec:alteration}

In order to overcome  the privacy issues of long-lived  IPv6 address prefixes, we propose to design and implement privacy-enhancing solutions, which provide unlinkable addresses. It is not the goal of this paper to provide a working solution. Instead we want to outline possible future research directions by sketching out three conceivable schemes with distinct properties.

\subsection{Scheme 1: Prefix Hopping}

\begin{figure}[htbp]
        \centering
                \includegraphics[scale=0.65]{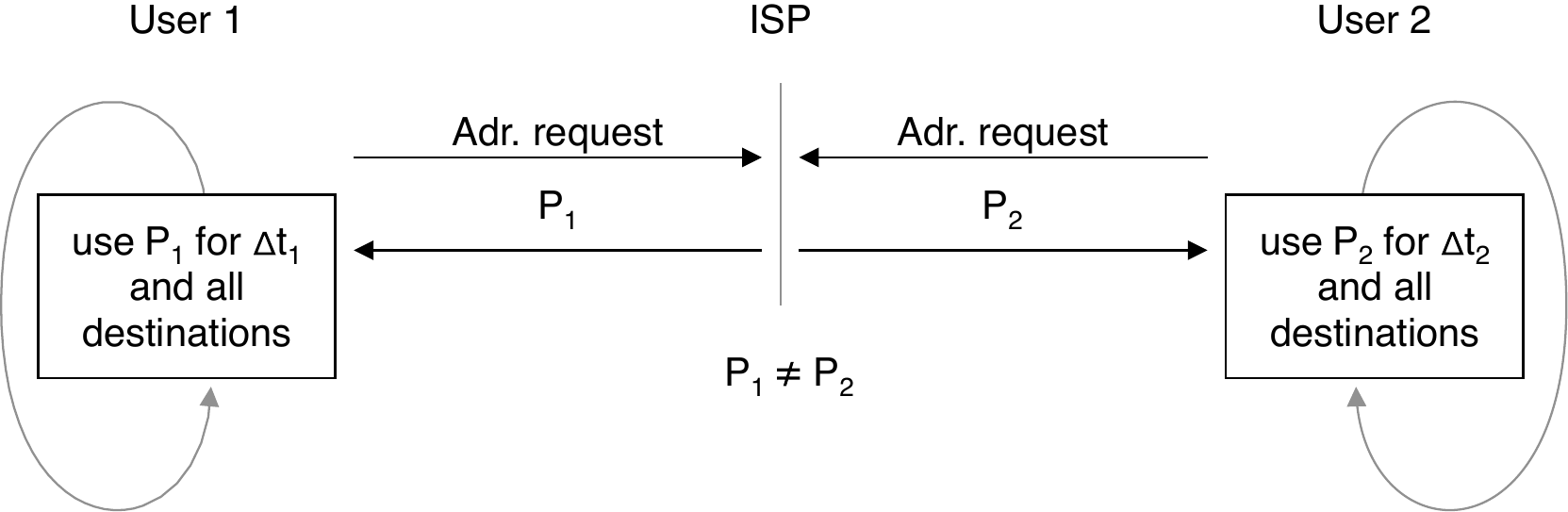}
        \caption{Prefix Hopping}
        \label{fig:fig_prefix_hopping}
\end{figure}

We assume that an adversary can link all activities of a user that are carried out using the same prefix, but cannot establish a link once the prefix of the user is changed (\emph{temporal unlinkability}). This basic level of unlinkability is already available today: Many ISPs terminate the broadband connection of residential customers in regular intervals, e.\,g., every 24 hours, and assign their customers a new IP address once the connection is established again. Prefix Hopping improves privacy by \textbf{changing prefixes} of customers more frequently, e.\,g., \textbf{within a few seconds or minutes}.  Consequently, for this scheme all outbound traffic of a customer's router uses a single prefix within a certain time frame (see Fig.~\ref{fig:fig_prefix_hopping}).

Switching to a new prefix should not be implemented by terminating the broadband connection, which would result in poor usability due to frequent service interruptions. Instead, seemless techniques such as short DHCP lease times in conjunction with deprecating previously assigned addresses should be considered. The level of privacy attainable by a certain value of the \emph{alteration frequency} should be studied in empirical evaluations using recorded traffic to determine reasonable values. Apart from this parameter there are several implementation options, whose impact on usability and privacy must be considered, such as:

\begin{enumerate}
	\item Do existing TCP connections have to be terminated (which impacts usability) once the prefix is replaced to guarantee unlinkability or can they be kept alive?
	\item Should the current prefix be announced by the customer's router on the internal network (which may cause issues on clients that cannot deal with frequent renumbering) or should it be hidden from the clients by employing network address translation techniques on the customers' router or at the ISP end?
\end{enumerate}

The advantage of Prefix Hopping is that it can be implemented without major changes to the existing protocols. On the other hand its protection is only limited due to the fact that all concurrent traffic remains linkable.

\subsection{Scheme 2: Prefix Bouquets}

\begin{figure}[htbp]
        \centering
                \includegraphics[scale=0.65]{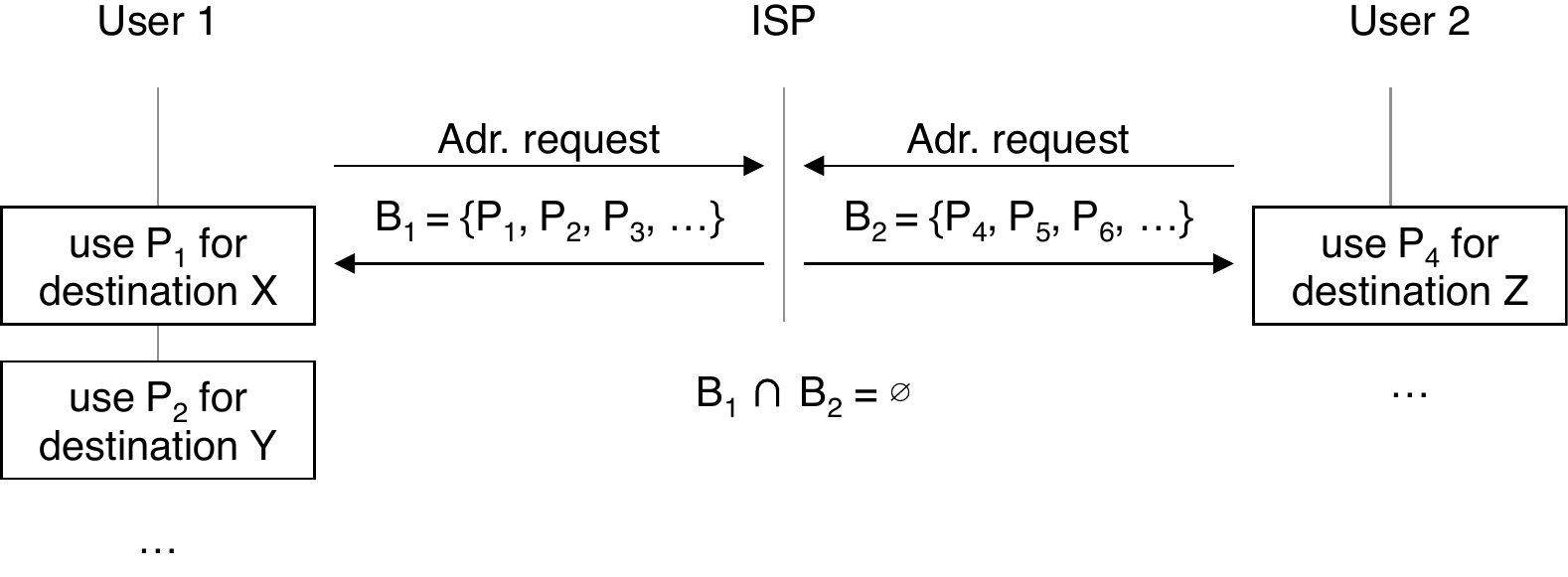}
        \caption{Prefix Bouquets}
        \label{fig:fig_prefix_bouquets}
\end{figure}

In this scheme the ISP does not only delegate one prefix to each customer at a given point in time, but a whole set of prefixes, a \emph{bouquet}, at once. This allows the customers to split their outbound traffic flows across a vast number of different prefixes so that even concurrent activities become unlinkable for adversaries (see Fig.~\ref{fig:fig_prefix_bouquets}). This scheme improves privacy by offering unlinkability on the \emph{transactional level}, i.\,e., on a per-connection (TCP) or even on a per-packet (UDP) basis.

Again, various implementations options with different usability and privacy properties are conceivable:

\begin{enumerate}
	\item One extreme case is to \textbf{use a different prefix for every outgoing flow}, i.\,e., every new TCP connection, every stream of related UDP packets, every single ICMP packet) regardless of its respective destination. This makes it impossible to link concurrent activities involving different destinations as well as consecutive requests that target the same destination. Neither web sites nor ad networks would be able to aggregate requests of their users based on the IP address any more. There may be compatibility issues, though: for security reasons some web applications may terminate HTTP sessions (maintained with cookies) if the source address changes between two consecutive requests.
	\item A more practical solution may be to \textbf{re-use prefixes on a per-destination basis}. This will still improve privacy significantly because there is no direct linkability on the IP layer between destinations if destinations collude. On the other hand, this variant cannot offer unlinkability against an ad network whose web server (which has a constant destination address) is accessed during visits of different web sites.
\end{enumerate}

With Prefix Bouquets ISPs can offer users unlinkability with various granularity, in principle even on a per-packet basis. This scheme cannot be implemented with currently available techniques and protocols, though. Even so, a transparent solution could be implemented using a network address translation service on the customer's router or at the ISP end.

\subsection{Scheme 3: Prefix Sharing}

\begin{figure}[htbp]
        \centering
                \includegraphics[scale=0.65]{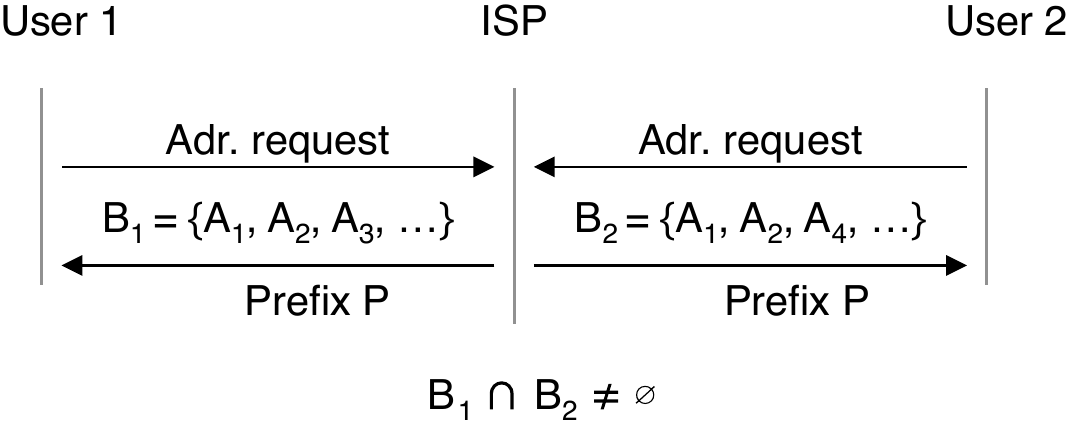}
        \caption{Prefix Sharing}
        \label{fig:fig_prefix_sharing}
\end{figure}

It can be expected that not all customers will need a ``whole'' IPv6 subnet, which would typically support on the order of $2^{64}$ hosts, at their disposal: They either do not have so many devices or they do not have to directly address so many network devices from the public Internet at all times. Prefix Sharing exploits this observation by \textbf{allocating a single prefix to multiple customers} that form an \emph{anonymity group}. Instead of delegating a unique prefix to each customer, the ISP now assigns individual (randomised) IPv6 addresses to its customers. As all IP addresses belong to the same IPv6 prefix, tracking individual users based on the prefix becomes impossible. Segregation of the  traffic of customers sharing the same prefix can be enforced via a firewall deployed within the ISP's router. There are different implementation options for Prefix Sharing:

\begin{enumerate}
    \item One option may be for the customer's router to \textbf{forward DHCP address requests} of network devices from the customer's network to the ISP. The DHCP server at the ISP will then assign the device an available IP address. While this variant can be implemented with off-the-shelf techniques, there may be privacy issues: due to the address requests the ISP may gain sensitive information about the customer's network and habits.
    \item Another option would be for the ISP to \textbf{pre-allocate a set} of random IPv6 addresses for each customer (see Fig.~\ref{fig:fig_prefix_sharing}) and distribute the whole set to the customers' routers once the broadband session is established. In this case IP address assignment would be carried out in the customer's router and remain hidden from the ISP. Distributing sets of addresses requires changes to the protocols running during session establishment and the software running on routers, though.
    \item Prefix Sharing could also be implemented without any changes on the customers' side via network address translation (NAT) at the ISP end.
\end{enumerate}

This scheme requires the ISP to keep track of individual addresses: On the one hand the ISP has to make sure that there are no address collisions at a given point in time, on the other hand from a privacy perspective it would be desirable that individual addresses are assigned to multiple customers over time.

With Prefix Sharing network nodes must still replace their address with a new one on a frequent basis in order to prevent tracking based on the interface identifier part of the address.

Coming from the theory of anonymity and unobservability, Prefix Sharing seems to be the canonical way to implement privacy on the IP(v6) level. Most practical solutions for anonymous communication make use of proxy-based solutions (such as virtual private networks, HTTP proxies and NAT) where multiple users share a single IP address, which usually causes performance and scalability issues. The scalability issue of Prefix Sharing is tied to the number of customers that can share a single prefix as well as the effort necessary to maintain the mapping table, in which the ISP stores the addresses that are assigned to a given customer. The latter of these two issues will be briefly discussed in Sect.~\ref{sec:implementation}. Therefore, Prefix Sharing may be seen as the last option if other solutions with better scalability are not usable.

\section{Implementation Considerations}
\label{sec:implementation}

In this section we will discuss requirements and considerations for the practical implementation of the aforementioned address alteration schemes.

\paragraph{Combination with Privacy Extensions.}

For prefix alteration to be effective it must always be used in conjunction with IPv6 Privacy Extensions \cite{RFC4941}. Whenever unlinkability is to be achieved by a different prefix, the interface identifier part of an IPv6 address has to be changed as well. As the required synchronization may be difficult to achieve in practice, integrating both techniques should be considered. The integration is straightforward in case the schemes are implemented using NAT techniques on the customer's router or by the ISP.

\paragraph{Control over Prefix Alteration.}

An important design decision is the question which party is to be in control of the prefix alteration process. We see two options here: 
\begin{itemize}
    \item The alteration can be fully delegated to the ISP, which could use network translation techniques to rewrite addresses accordingly. As customers are already willing to trust their ISP with their traffic, this delegation does not introduce new privacy issues. We assume that ISPs that actually do deploy prefix alteration techniques are going to support their customers, rather than undermine their efforts to establish unlinkability against third parties. The delegation may be advantageous for implementation reasons because customer premises equipment would remain unmodified. 
    \item A fundamentally different option  is to carry out the \textbf{alteration in the customer's router}. In this case the ISP would have to offer the router a set of prefixes once the connection is established for the router's disposal. This option requires changes to the software on customer premises equipment as well as on the ISP end, though.
\end{itemize}

\paragraph{Generating and Storing Collision-free Random Sets of Addresses.}

Multiple prefixes (Scheme 1 and 2) or addresses (Scheme 3) assigned to one customer over time must not be linkable by an adversary, i.\,e., addresses allocated to a customer must not span a closed range or share all a common (shorter) prefix. Therefore, ISPs need techniques to generate  a collision-free series of random addresses for each customer.

A naive approach would be to store pre-generated sets of addresses for each customer in the ISP's routers. This could be infeasible due to poor scalability, though. Another option is to implement a network address translation scheme that substitutes addresses on the fly with a \emph{tweakable cipher} as proposed in \cite{Enlisting-ISPs}.

We propose to also take time-memory-tradeoff techniques such as rainbow tables (cf.~\cite{Oechslin2011}) into consideration for this purpose. This approach employs a cryptographic hash function as well as a so-called reduction function, which maps hash values back to random IP addresses. The ISP can generate a list of random IP addresses for a customer by starting out with a random address. A random chain of addresses is then obtained by applying the hash function and the reduction function in alternating manner a fixed number of times. Only the last hash value and the first address of a chain have to be stored in the routing table. In order to determine for which customer an incoming packet (having a random address) is destined, the ISP looks up the address in the rainbow table. For a detailed description of the technique we refer the reader to \cite{Oechslin2011}.

In a number of countries ISPs are obliged to store the mapping between issued addresses and customers for an extended period of time (\emph{data retention}) in order to be able to answer enquiries by law enforcement agencies. Tweakable ciphers or rainbow tables may help to reduce the possibly demanding storage requirements caused by the large numbers of prefixes or addresses assigned to each customer.

\section{Limitations}
\label{sec:limitations}

On first sight prefix alteration techniques conflict with end-to-end reachability of nodes. After all, providing services that should be reachable from the public network (incoming connections) requires static and possibly life-long constant prefixes.

As proposed in \cite{Enlisting-ISPs} end-to-end reachability could be maintained by allocating an additional static prefix for each customer. Incoming connections to the static address can either be directly routed to the customer's network by the ISP or mapped to the customer's dynamic prefix using NAT techniques.

Even though address alteration schemes may improve privacy, they cannot protect against attacks on the higher levels of a communication network. For instance, let's consider Prefix Hopping (Scheme 1). Tabbed browsing and client-side JavaScripts can be used to vanquish unlinkability. Imagine a tracking ad network that displays a banner ad in one tab and re-downloads it periodically with a client-side script (e.\,g., every minute). If the banner image resides at a unique URL, i.\,e., each user retrieves a different URL,  the ad network will be able to follow along the altered prefixes, rendering Prefix Hopping useless as long as the respective browser tab is not closed.

Tracking across multiple protocol stack layers is a fundamental issue, though, which can only be resolved if all identifiers on all layers are switched simultaneously \cite{Lindqvist:2008:PPP:1456403.1456416}. Therefore, prefix alteration cannot cure tracking issues that exist in HTTP. Its purpose is rather to make sure that IPv6 does not introduce new tracking possibilities that will have to be overcome in higher protocol layers with solutions such as overlay networks (e.\,g., Tor).

\section{Conclusions}

IPv6 prefix alteration schemes would help to improve the protection of residential Internet connections compared to the current state in IPv4. 

In the current discussion of the ``long-lived prefix issue'' privacy advocates limit themselves to call for maintaining the status quo, i.\,e., issuing a new prefix upon each manual or daily reconnect of the customer's router. We believe that the privacy community must strive for more: The upcoming introduction of IPv6 is a very seldom opportunity to \emph{improve} the overall level of privacy for many users.

However, IPv6 prefix alteration should not be considered as a universal privacy solution for all applications on the Internet. As long as technologies such as HTTP cookies are enabled on higher protocol levels, tracking is obviously still possible. Anyway, if privacy on the lower levels of the protocol stack is ensured, this provides new opportunities for ensuring privacy in the whole communication network. Future work should look into such issues and evaluate different alteration schemes in real-world settings. 

\subsection*{Acknowledgements}

The authors are grateful to Christoph Gerber and Karl-Peter Fuchs for their insightful comments and suggestions.

\label{sect:bib}
\bibliographystyle{plain}

\bibliography{HeAF2012}

\end{document}